# Design and analysis of MoTe$_2$-based efficient photonic devices for the solar cell and photodetector applications

Md. Naeemur Rahman[1], Md. Alamin Hossain Pappu[1], Md. Islahur Rahman Ebon[1,2], Abdul Kuddus[3], Dinesh Pathak[4], and Jaker Hossain[1*]

[1]*Photonics & Advanced Materials Laboratory, Department of Electrical and Electronic Engineering, University of Rajshahi, Rajshahi 6205, Bangladesh.*
[2]*Department of Electrical and Electronic Engineering, Gono Bishwabidyalay, Savar, Dhaka 1344, Bangladesh.*
[3]*Ritsumeikan Global Innovation Research Organization, Ritsumeikan University, Shiga 525-8577, Japan.*
[4]*Department of Physics, The University of the West Indies, St. Augustine Campus, Trinidad and Tobago.*

**Abstract**

A systematic survey and subsequent research have been made on MoTe$_2$-based *n*-CdS/*p*-MoTe$_2$/*p$^+$*-CGS device in solar cell and photodetector field. The optimization has been established by altering the various properties of each constituent layer through numerical computation. The performance of the MoTe$_2$ photonic device has been probed with and without CGS back surface field (BSF) layer in details. The proposed *n*-CdS/*p*-MoTe$_2$/*p$^+$*-CGS photonic device exhibits markedly improved cell efficiency, $\eta$ of 32.92 % with V$_{OC}$ of 0.97 V, J$_{SC}$ of 41.21 mA/cm$^2$, FF of 82.73% and responsivity, R of 0.74 A/W as well as detectivity, D$^*$ of 2.36×10$^{16}$ Jones at a wavelength of 1000 nm. These simulation outcomes reveal the strong potential of MoTe$_2$ absorber along with the novel and improved structure for highly-efficient solar cells and photosensors that capable of high detection capability.

**Keywords:** MoTe$_2$, CdS, CGS, solar cell, photodetector, SCAPS-1D.

## 1. Introduction

Renewable technologies are one of the most possible ways to mitigate the huge future energy crisis offering green energy sources. Photovoltaic (PV) and photosensors (PS) technologies have proved itself as one of the promising, reliable, and economical sources among the renewable technologies. PV cells convert sunlight into electricity where photosensors covert optical signal into electrical signals [1-2]. Thin film solar cells; especially organic, inorganic, perovskites are becoming emerging due to their notable performance and/or low production cost compared to silicon-based PV cells, which conventionally involve high temperatures and expensive fabrication system [3-4]. The PV efficiency of 20-22% has been obtained already by inorganic thin film solar cells with CdTe, CuInGa(S, Se) semiconductors [5-6]. The large number of works (both of theoretical and experimental) have been performed by the researchers on ternary chalcopyrite compounds-based thin film solar cells, such as CdTe, CdS, $CuInSe_2$, $CuInGaSe_2$, $Cu_2ZnSn(S, Se)_4$, $AgInSe_2$, $CuInTe_2$, $Cu(In, Ga)Te_2$, $BaSi_2$, SnS, $Cu_2SnS_3$, $CuSbS_2$, $CuSbSe_2$, GeSe, $FeS_2$, $FeSi_2$, and SnSe, $Sb_2S_3$, $Sb_2Se_3$, $WS_2$, $WSe_2$, $MoS_2$, $MoSe_2$, $MoTe_2$ transition metal dichalcogenides (TMDCs) [7]. A less amount of work has been done as a photosensors compared to solar cells. $AlO_x$, Si and GaAs are the materials which are frequently used in photdetector filed. These photosensors provides a responsivity and detectivity in the range of 0.26-0.63 A/W and $4.3 \times 10^{11}$-$1.10 \times 10^{13}$ Jones [8-10]. On the contrary, a few amounts of work have been performed as a solar cell and photosensors in the same structure. $CH_3NH_3PbI_{3-x}Cl_x$ perovskite compound provides an efficiency close to 23% where responsivity (R) of 0.45 A/W, and detectivity ($D^*$) of $6.01 \times 10^{11}$ Jones. But Perovskite compound contains a several drawbacks such as defenseless against moisture, unstable at high temperatures [11-12]. Solar cells and photosensors performance have been also performed in the same structure by using organic and inorganic compound such as O6T-4F, PTB7-Th, and $Sb_2S_3$. But they provide low performance parameters such as efficiency (1.17-13.23%), responsivity (0.35-0.52 A/W) and as well as detectivity ($1.13 \times 10^{13}$-$1.10 \times 10^{14}$ Jones) [13-14]. Another research has been done on $CuTlS_2$ that conducts on both solar cell and photodetector with a responsivity of 0.67 A/W and a detectivity of $9.54 \times 10^{16}$ Jones [15]. Among these TMDCs, $MoTe_2$ is used in this heterojunction PV cell as an absorber. The $MoTe_2$- based some solar cells have been investigated earlier. But the novel structure of $MoTe_2$ with the CdS window and CGS BSF have not been introduced yet. Also, the analysis of both of the solar cell and photosensor (PS) has not also been studied earlier.

$MoTe_2$ is a layered element display narrowed photoluminescence response with no bound exciton shoulders including high carrier mobility, low defects density and impressive optical properties [16-19]. TMDCs- and their alloys-based heterostructures allow tuning and tailoring of the band structures within the stoichiometric composition limit and explore characteristic associated with the interlayer coupling. Molybdenum Telluride ($MoTe_2$) is an emerging layered semiconducting material comprising distinct structural, electronic, and photonic properties including low charge transfer resistance, high thermal stability, and favorable bandgap, $E_g$ of 1.1 eV which is almost equal to that of Si (1.0 eV). It has rarely been identified for designing a primitive unit of complex structures [20]. $MoTe_2$ and Si contains some superb electrical and optical properties but Si-based solar cells have a variety of drawbacks, including cost, weather dependence, space requirements, pollution concerns, rigidity, and highly expensive for the case of solar energy storage, and so on [21]. Bulk $MoTe_2$ composed of hexagonal sheets of Mo are sandwiched between trigonal-prismatic coordination of two Te atoms hexagonal planes as Te–Mo–Te units with Fermi arcs on the film surface, diverse magneto-transport characteristics, anomalous Hall effect, pressure-driven superconductivity and distinct quantum transport owing to chiral anomaly [22-23]. $MoTe_2$ is unambiguously classified as orthorhombic (a=3.458 Å, b=6.304 Å, c=13.859 Å) crystal structure [24]. $MoTe_2$ also exhibits phonon-limited high mobility at room-temperature [24-25]. Experimental reports reveal the long stability of the $MoTe_2$ sample after synthesis. Additionally, synthesized $MoTe_2$ has low charge transfer resistance and high electrocatalytic properties at the electrolyte-electrode interface. It also has high optical absorption, high charge carrier mobility, good electrical conductivity and thickness-dependent physical and chemical properties [20, 26-29]. With a tunable low band gap near Si, $MoTe_2$ has been found as a strong semiconductor to be used as a non-toxic, economic, abundant in nature and stable which can be used as absorber and BSF layer in solar cells (SCs). Its crystals contain the characteristics of Van der Waals formation with less interface defect concentration [29]. $MoTe_2$ is also promising to be applied as counter electrodes in dye-sensitized (DS) SCs [19, 26, 29-30].

In addition, CdS is widely researched as a prominent n-type semiconducting material owing to its tunable wide optical bandgap, elevated carrier density, and excellent transmission properties, coupled with remarkable stability even under prolonged exposure to light [31]. Also, CdS belongs to group II-VI and exhibits two phases, namely cubic (a=5.832 Å) and hexagonal (a=4.130 Å, c=6.703 Å) [21]. On the other hand, $CuGaSe_2$ (CGS) which is potential semiconductor to be used as BSF layer, conventionally deposited by molecular beam epitaxy (MBE), chemical vapor transport (CVT), spray pyrolysis, metal-organic chemical vapor deposition (MOCVD), or electrodeposition methods are found as potential BSF layer with large $E_g$ and stability [32]. The crystal structure of CuGaSe$_2$ can be confidently identified as tetragonal in which a=5.596 Å and c=11.003 Å [27, 33]. The CGS shows a tunable optical gap of ~1.66 eV with a higher coefficient of absorption, α of ~ $10^4$–$10^5$ cm$^{-1}$ in the insight to longer wavelengths regions, which leads to be strong potential candidate as BSF layer for obtaining high efficiency in the proposed heterojunction photovoltaic [34]. The higher band gap energy of CGS allows significant absorption of longer wavelength photons of irradiated spectrum.

In the present work, the investigation has done on the potentiality of MoTe$_2$-based $n$-CdS/$p$-MoTe$_2$/$p^+$-CGS heterostructure device as solar cell and photosensor. By varying layer width, carrier density, volume and interface defect density of used semiconductors such as CdS window, MoTe$_2$ absorber, and $p^+$-CGS BSF layers, the optimum condition is determined. These systematic studies provide resources for the MoTe$_2$ -TMDCs-based high-performance photonic (solar cell and photosensor) devices.

## 2. Methodology and Simulation parameters

Simulators aid in analyzing interested cell structures and predict the impact of layer parameters on device performance. Recent years have witnessed significant efforts performed in developing sustainable energy systems through state-of-the-art solar cell simulators. A succinct evaluation of solar cell simulators is sometimes required to identify the most reliable tool for assessing and anticipating photovoltaic performance. The simulation software plays a crucial role in designing low-cost, environment-friendly solar energy systems and determining potential PV system outcomes for various projects, resulting in a continual advancement in the solar energy technology landscape. Figure 1(a) and (b) exhibit the structure and band schematic of the proposed $n$-CdS/p-MoTe$_2$/$p^+$-CGS configuration. The window and BSF layers are $n$-type cadmium sulfide (CdS) and $p^+$-type copper gallium selenide (CGS) layers with bandgaps of 2.4 and 1.66 eV, respectively. Almost zero conduction band offset (CBO) was determined at $n$-CdS/p-MoTe$_2$ interface while the offset of valence band (VBO) of 0.1 eV at MoTe$_2$/$p^+$-CGS. Front and rear metal contacts are aluminum (Al) and nickel (Ni) with work function (WF) of 4.2 eV and 5.2 eV, in turn. These contacts offer low resistive path for the successive carrier (electron or hole) transportation from active layers to outer circuits.

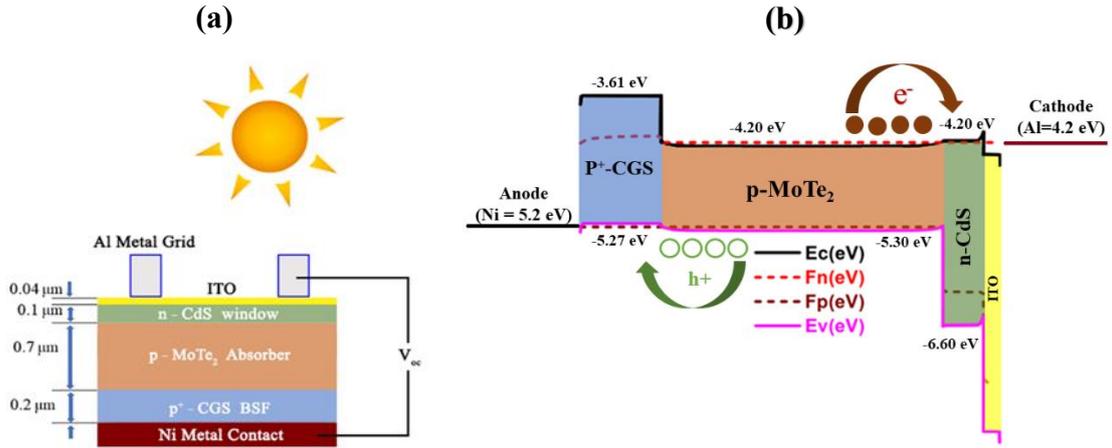

**Figure 1:** (a) The schematic diagram of the proposed *n*-CdS/*p*-MoTe$_2$/*p$^+$*-CGS double-junction solar cell, and (b) band alignment of the corresponding solar cell structure.

The MoTe$_2$-based n-CdS/p-MoTe$_2$/p$^+$-CGS photovoltaic structure has been delved embarking the solar cell capacitance simulator which is one-dimensional (SCAPS-1D) that solves Poisson and carrier continuity equations (Equations 1-3) related to electrons and holes for computing solar cell characteristics like current voltage J-V, capacitance voltage C-V, capacitance-frequency C-f and quantum efficiency (QE) [31].

$$\frac{\partial^2 \Psi}{\partial x^2} + \frac{q}{\varepsilon}[p(x) - n(x) + N_D - N_A + \rho_p + \rho_n = 0] \quad (1)$$

(Hole continuity equation) $\frac{1}{q}\frac{\partial J_p}{\partial x} = G_{op} - R(x)] \quad (2)$

(Electron continuity equation) $\frac{1}{q}\frac{\partial J_n}{\partial x} = -G_{op} + R(x)] \quad (3)$

where, N$_A$ = volume of acceptors, N$_D$ = volume of donors, $\varepsilon$ = dielectric coefficient, $q$ = the charge of electron, J$_p$ = density of hole current and J$_n$ = density of electron current. The electrostatic potential is denoted by the symbol $\Psi$, G$_{OP}$ is the whole rate of carrier generation and R is the rate of recombination, p, n is the free hole and electron volume, $\rho_p$ is for the allocation of hole and $\rho_n$ for electron. So far, carriers transport properties in the layers are determined by drift-diffusion Equations (4, 5) [31].

$$J_p = -\frac{\mu_P p}{q}\frac{\partial E_{Fp}}{\partial x} \quad (4)$$

$$J_n = -\frac{\mu_n n}{q}\frac{\partial E_{Fn}}{\partial x} \quad (5)$$

Where, the hole and electron mobilities are denoted by $\mu_p$ and $\mu_n$, E$_{Fp}$ and E$_{Fn}$ are the sign of Fermi levels of hole and electron charge carriers, in sequence.

To evaluate the gadget's execution and sensibility as a photosensor (PS), basically two key criteria are crucial: Responsivity (R), expressed in A/W as well as Detectivity ($D^*$) which is measured in Jones. The following equations showcased how they have determined [35].

$$R = \frac{q\eta\lambda}{hc} \qquad (6)$$

$$D^* = \frac{R}{\sqrt{2qJ_0}} \qquad (7)$$

Here, R, q, η, h, c, λ and $J_0$ represent responsivity, electron charge, quantum efficiency (QE), Planck constant, wavelength and the reverse saturation current, respectively.

The solar PV device was illumined under the solar spectrum of AM 1.5G with $P_W$ =100 mW/cm². Fixed electron and hole thermal velocity of $1.0\times 10^7$ cm/s and operating temperature (WF) of 300 K were considered considering practically obtained values. The optical model for each layer used default optical data provided by the SCAPS-1D simulator. The interface defects also play vital impact on PV efficiency; therefore, an interface defect density of $10^{11}$ cm$^{-2}$ was introduced at CdS/MoTe$_2$ and MoTe$_2$/CGS interfaces. The radiative, Shockley-Read-Hall (SRH), and Augur recombination were taken into account with recombination coefficients of $2\times10^{-9}$, $1\times10^{-14}$ and $1\times10^{-29}$ (cm³/s), respectively. Simulation parameters of each photoactive material were collected from literatures [19, 31, 36] and summarized in Table 1.

**Table 1.** Different input characteristics that are utilized for active layers in this solar cell.

| Parameter | *n*-CdS | *p*-MoTe$_2$ | $p^+$-CGS |
|---|---|---|---|
| Width (nm) | 100 | 700 | 200 |
| Optical Gap (eV) | 2.4 | 1.1 | 1.66 |
| The electron affinity | 4.2 | 4.2 | 3.61 |
| Relative dielectric permittivity | 10 | 13 | 10.9 |
| CB effective DOS (cm$^{-3}$) | $2.2\times 10^{18}$ | $1.0\times10^{15}$ | $2.2\times10^{17}$ |
| VB effective DOS (cm$^{-3}$) | $1.8\times10^{19}$ | $1.0\times10^{17}$ | $1.8\times10^{18}$ |
| Electrons Mobility (cm²/Vs) | $1.0\times10^2$ | $1.1\times10^2$ | $1.0\times10^2$ |
| Holes mobility (cm²/Vs) | $2.5\times10^1$ | $4.26\times10^2$ | $2.5\times10^2$ |
| Donor volume, $N_D$ (cm$^{-3}$) | $1.0\times10^{18}$ | 0 | 0 |

| Acceptor volume, $N_A$ (cm$^{-3}$) | 0 | $1.0\times10^{15}$ | $1.0\times10^{18}$ |
| --- | --- | --- | --- |
| Defects (cm$^{-3}$) | $1.0\times10^{15}$ | $1.0\times10^{15}$ | $1.0\times10^{15}$ |
| Radiative Coefficient (cm3/s) | 0 | $1.0\times10^{-9}$ | 0 |

## 3. Results and discussion
### 3.1. Device performance with MoTe$_2$ semiconductor
#### 3.1.1. Device as Solar cell

Figure 2 exhibits the consequences of MoTe$_2$ width on the PV execution of *n*-CdS/*p*-MoTe$_2$/*p*$^+$-CGS device at a layer span of 100-1300 nm, carrier doping of $10^{13}$-$10^{19}$ cm$^{-3}$ and the defect density of $10^{12}$-$10^{18}$ cm$^{-3}$. The PCE increased from ~20.47 % to 32.67% with the J$_{SC}$ from ~25.49 to 42.34 mA/cm$^2$ and FF 77.77% to 82.27%, almost logarithmically, and reached saturation while V$_{OC}$ decreased from ~1.03 to 0.93 V sub-linearly with enhancing the MoTe$_2$'s width from 100 to 1300 nm in Figure 2(a). The V$_{OC}$ of the absorber thickness is down to 0.95 V at 900 nm and reaches constant regardless of the further increase in absorber thickness. An increased absorber thickness absorbs higher incident photons and generates higher photocurrent. On the contrary, the larger diffusion length at higher thickness causes the noticeable recombination, resulting in, reduced V$_{OC}$. The lack of adequate effective area hinders efficient photon absorption, primarily because the back metal contact is situated in close proximity to the depletion region within a thinner absorber layer [37]. Thus, the PCE of around 32.92%, J$_{SC}$ of 41.2 mA/cm$^2$, and FF of 82.73% obtained at optimized absorber layer breadth of 700 nm, carrier concentration of $10^{15}$ cm$^{-3}$, and defects of $10^{15}$ cm$^{-3}$, respectively, thinking the tradeoff among PV parameters.

Figure 2(b) depicts the performance parameters of the absorber layer at varying doping concentrations, N$_A$ of MoTe$_2$ of $10^{13}$-$10^{19}$ cm$^{-3}$ at a secured layer width of 700 nm, and defects of $10^{15}$ cm$^{-3}$. With increasing the MoTe$_2$ doping from $10^{13}$ to $10^{14}$ cm$^{-3}$, the V$_{OC}$ is almost unchanged, however, increases quickly from 0.96 to 1.04 V exponentially when N$_A$ changes from $10^{14}$ to $10^{19}$ cm$^{-3}$. The elevated V$_{OC}$ is attained as a consequence of the heightened built-in potential established in the interface of the absorber layer, stemming from a higher doping concentration [7]. The FF increases from 82.46 to 86.68%, eventually, the PCE rises from 32.69% to 36.23% with increasing N$_A$ from $10^{13}$ to $10^{17}$ cm$^{-3}$, one by one. The worth of J$_{SC}$ is found mostly stable at 42.14 - 41.31 mA/cm$^2$. The increment in doping volume enhances the V$_{bi}$ which is known as built-in potential initially, while further increase causes visible recombination as indicated by sudden zigzag behavior of FF beyond $N_A$ of $10^{17}$ cm$^{-3}$. This nature of FF occurs because of the diode ideality factor [31]. Thus, the $N_A$ of $10^{15}$ cm$^{-3}$ was determined as an adjusted doping level in the MoTe$_2$ absorber respecting the change in PV parameters.

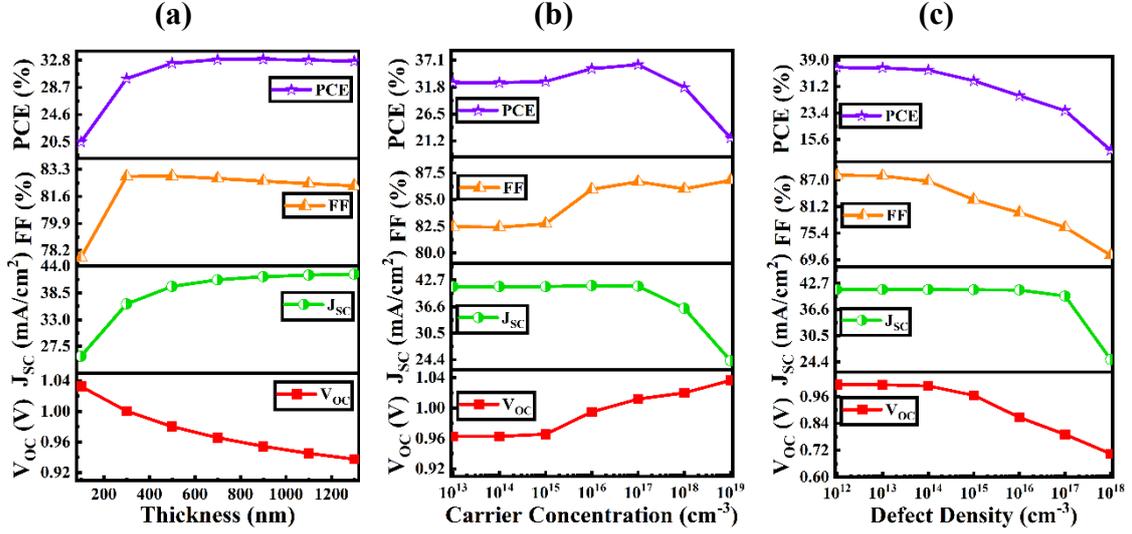

**Figure 2:** The variation of PV performance with MoTe$_2$ section of $n$-CdS/$p$-MoTe$_2$/$p^+$-CGS device at a different (a) layer thickness; (b) acceptor concentration, $N_A$; (c) defect density, $N_t$.

Basically, the impurity density for different layers is selected from default settings of SCAPS-1D and then altered at a range to determine the optimized defect density for the proposed PV cell structure. Figure 2(c) demonstrates the performance parameters of the absorbing layer concerning the defects, $N_t$ of MoTe$_2$ from $10^{12}$ to $10^{18}$ cm$^{-3}$ at a fixed width of 700 nm and doping amount of $10^{15}$ cm$^{-3}$. With increased defect concentration, $V_{OC}$ mitigates since 1.01 to 0.70 V drastically with changing the $N_t$ from $10^{12}$ to $10^{18}$ cm$^{-3}$. Similarly, the short circuit current, FF and PCE decrease from 42.22 mA/cm$^2$ to 24.94 mA/cm$^2$, 88.07% to 70.5% and 36.8% to 12.9%. At a higher value of defect density, the recombination due to Shockley-Read-Hall (SRH) ($\Re_{SRH}$) is dominating, resulting in a rapid diminishing in PV parameters is visualized [31].

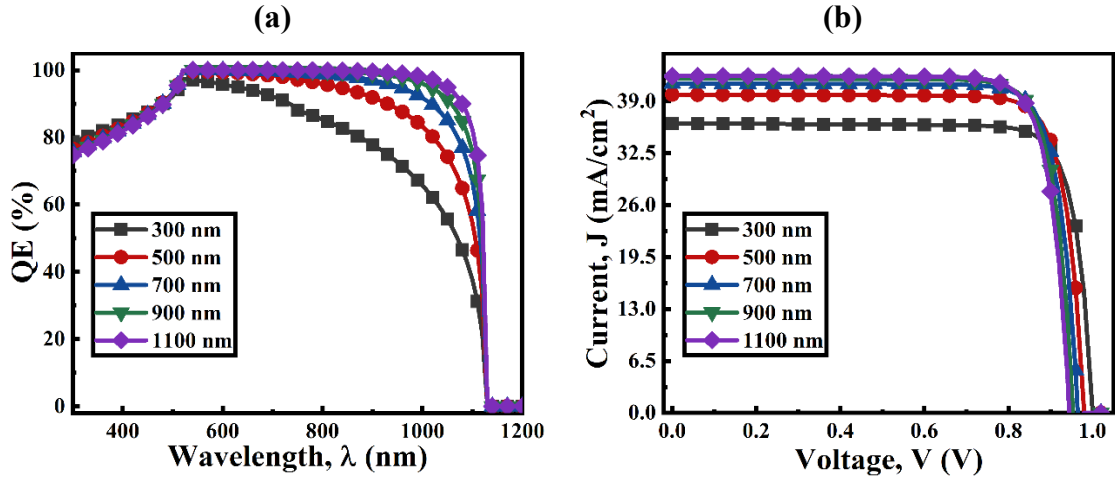

**Figure 3:** (a) The quantum efficiency (QE) and (b) corresponding J-V characteristics at a different thickness of MoTe$_2$ absorber layer with *n*-CdS/*p*-MoTe$_2$/*p$^+$*-CGS heterostructure.

Figure 3(a) and (b) shows the quantum efficiency (QE) and J-V characteristics curve of *n*-CdS/*p*-MoTe$_2$/*p$^+$*-CGS at different thickness layer of MoTe$_2$ ranges of 300-1100 nm, respectively. The carrier's ratio is created by absorbed photons which are fortuitous with wavelength (λ) on the device to incident photon energy is reasonably determined by QE (%) [26, 38]. At thin layer thickness of ≤300 nm, an absorption of incident solar spectrum was found insufficient with QE from 65.7 to 97% to that originate relatively smaller current (25.49 mA/cm$^2$) within a wavelength range of 520-1100 nm. With significant MoTe$_2$ absorber layer thickness of ≥800 nm, the QE is found from 97% -100% over a wide region at 520-1000 nm. As a result, the markedly improved J of ∼42.16 mA/cm$^2$ and thereby the PCE exceeding 32%. The reduction in voltage from 0.94 to 1.03 V owing to the photogenerated carrier recombination with a longer diffusion length [38].

### 3.1.1. MoTe$_2$ as a photosensor device

Figure 4 delves into the preface of MoTe$_2$, the photon-absorbing layer, thickness, doping, and defects on the performance on the MoTe$_2$ PD. Specifically, exploration of varying the width, doping volume and defects density of the MoTe$_2$ layer from 300 nm to 1300 nm, from 10$^{13}$ cm$^{-3}$ to 10$^{19}$ cm$^{-3}$ and from 10$^{11}$ cm$^{-3}$ to 10$^{17}$ cm$^{-3}$, respectively influences device behavior as depicted in the figure.

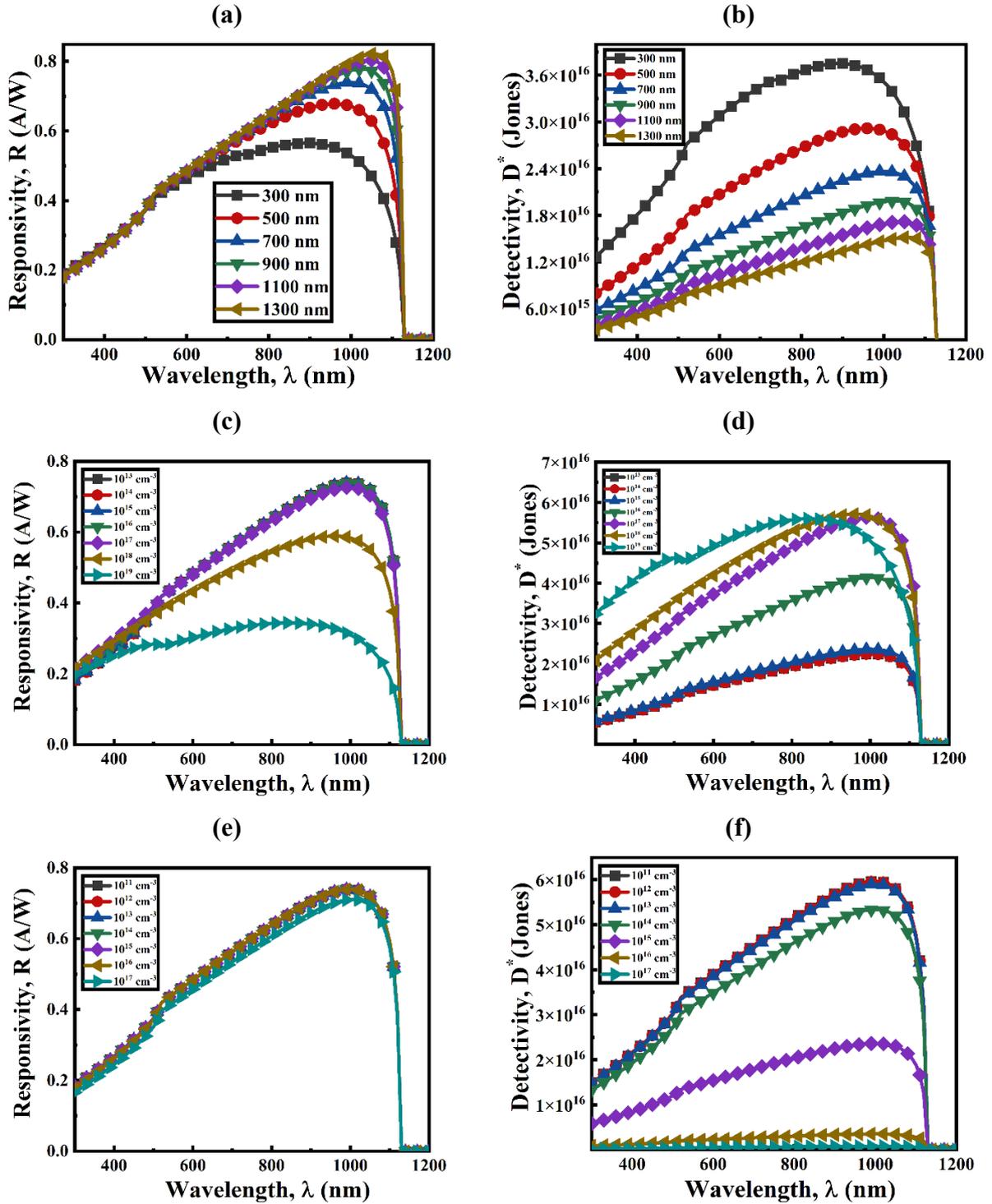

**Figure 4:** (a) Responsivity and (b) detectivity on different thickness; (c) responsivity and (d) detectivity on different doping density; (e) responsivity and (d) detectivity on different defects of MoTe$_2$ absorber layer with *n*-CdS/*p*-MoTe$_2$/*p*$^+$-CGS heterostructure PD.

Figure 4(a) and 4(b) adorn the variations in R and D$^*$ of the photosensor (PS) as a function of wavelength, with different thicknesses of the MoTe$_2$ layer taken into account. Increasing the thickness of the absorbing layer from 300 nm to 1300 nm results in a step-by-step elevation of each R as well as $D^*$. This upshot is especially significant at wavelengths near the optical boundary of the MoTe$_2$ absorbing layer, due to the critical role of photon energies near the bandgap in effective absorption, leading to higher photocurrent generation [39]. The values of R and $D^*$ are largely dependent on the quantum efficiency (QE), which explains the observed changes in R and $D^*$ at a wavelength of 1000 nm in these figures. During the variation of width, the PS unveils an incremental responsivity from 0.57 A/W to 0.82 A/W whereas detectivity slightly decrease from $3.75 \times 10^{16}$ Jones to $1.5 \times 10^{16}$ Jones. But at an optimized thickness of 700 nm for the absorbing layer, the PS achieves maximum responsivity and detectivity standard of 0.74 A/W and $2.36 \times 10^{16}$ Jones, in that order. This peak execution is attained at a wavelength of 1000 nm, highlighting the applicability of this photosensor in the near-infrared (NIR) region.

In Figure 4(c) as well as (d), the responsiveness and detection capability consistently demonstrate specific behavior across various doping concentrations. Notably, the responsivity is slightly decrease from 0.74 A/W to 0.34 A/W whereas detectivity is slightly rise from $2.36 \times 10^{16}$ Jones to $5.6 \times 10^{16}$ at 1000 nm wavelength with the dopant volume of $10^{16}$ cm$^{-3}$.

Figures 4(e) and (f) illustrate the R and $D^*$ of this PS device across varying defect concentrations within the absorber, as previously discussed. Both of these are crucial in determining the photo-detection capabilities of this MoTe$_2$ heterostructure photonic device. For defect concentrations varies from $10^{11}$ to $10^{17}$ cm$^{-3}$, the grades of responsivity and detectivity decreasing from 0.74 to 0.70 A/W and $5.97 \times 10^{16}$ to $1.16 \times 10^{14}$ Jones, one by one at a wavelength of 100 nm. But, an increase in defects beyond this threshold results in a notable decline in both. The decline in photosensor performance, particularly in detecting the NIR region, can be traced back to an elevated rate of recombination occurring at localized energy states due to increased defect concentrations [40].

### 3.2. Device performance with CdS window

The window layer having a thin thickness, large bandgap, and the least series resistance sandwiched in a heterojunction TFSC is a major criterion to develop a p-n junction with an absorber layer to achieve high optical throughput. Figure 5(a) demonstrates the variation of PV parameters with the change in thickness from 50 to 350 nm of the *n*-CdS window layer. Herein, there slight impact on the $V_{OC}$, $J_{SC}$, and FF as thickness varies. The $V_{OC}$ slightly decrease from 0.97 to 0.96 V, the $J_{SC}$ of 41.65 to 39.5 mA/cm$^2$, the FF of 82.81 to 82.68% and PCE of 33.32 to 31.48%. However, the photons are absorbed outside of the space charge width when the film's thickness increases beyond 300 nm, raising the minority charge current noticeably. As a result, the efficiency decreases as the charge recombination rate starts to dominate [41]. Thus, the power of the window layer thickness does not considerably affect PV parameters on the performances of this solar cell device when the width is situated in the span of several tenths to a hundred nanometers.

The alteration in the standards of PV parameters have an increment in the doping concentration of the window layer n-CdS varying from $N_D$ of $10^{15}$ to $10^{21}$ cm$^{-3}$ is displayed in Figure 5(b). The almost constant voltage and current are detected at 0.97 V whereas $J_{SC}$, and PCE decrease to 40.36 mA/cm$^2$ and 32.73%. Additionally, FF shows a zigzag nature in the range of $10^{17}$ to $10^{19}$ cm$^{-3}$ doping concentration. This kind of change is happened because of the diode ideality factor [31]. Therefore, the moderate PCE is found at ~32.92% at $N_D$ of $10^{18}$ cm$^{-3}$, which was started to decrease for further increase in $N_D$. Thus, the $N_D$ of $10^{18}$ cm$^{-3}$ in the CdS layer has adjusted as the chosen optimal value considering the sufficient photon passing through the window to the absorber layer and the least carrier recombination for this heterojunction cell.

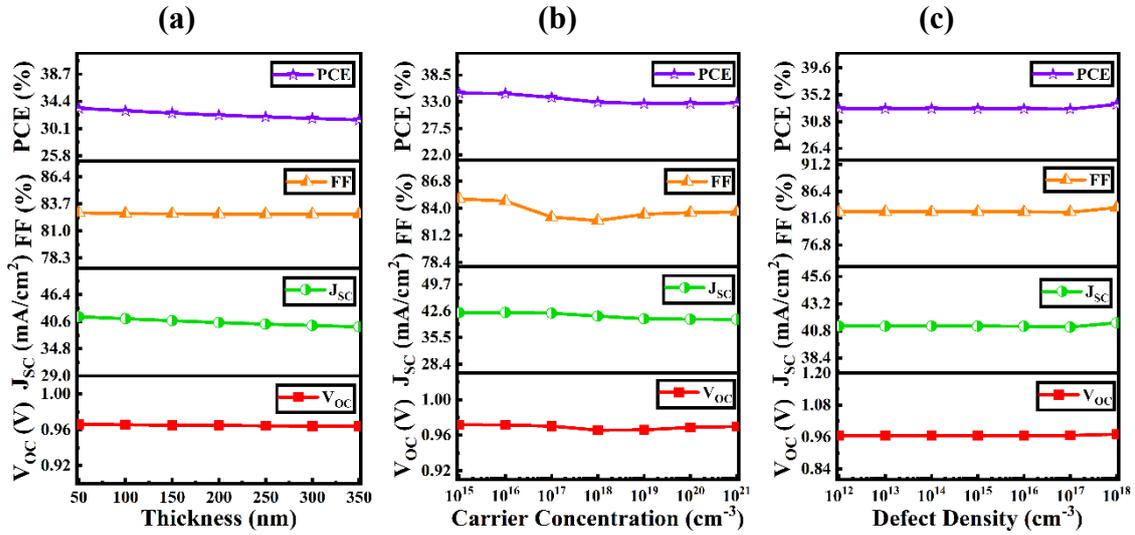

**Figure 5:** The alteration of PV performance with CdS layer (a) width; (b) carrier; (c) bulk defects of *n*-CdS/*p*-MoTe$_2$/*p*$^+$-CGS solar device.

Figure 5(c) shows the alteration in PV characteristics of designed cells with increasing in bulk defects, $N_t$ in the n-CdS window layer in the span of $10^{12}$-$10^{18}$ cm$^{-3}$. The $V_{OC}$ is stable and almost constant at 41.2 mA/cm$^2$, whereas $J_{SC}$, FF and PCE are slightly rise from 41.20 to 41.5 mA/cm$^2$. 82.73 % to 83.48% and 32.92% to 33.62%, respectively. Since, defects in photoactive materials act as recombination centers for photogenerated carriers, the PV parameters are markedly impacted by the defect density as SRH recombination in window regions as indicated by severe drop in FF. Besides, an inclusion defect in the CdS window, the dark current might increase drastically [38]. Thus, the bulk defect density of window layer $N_t$ of $10^{15}$ cm$^{-3}$ is found as an optimized value, which is in the range of realistic values obtained in practical devices.

### 3.3 Role of CGS back surface layer
### 3.3.1 MoTe$_2$ Solar cell

Figure 6 depicts the change in PV parameters of ITO/*n*-CdS/*p*-MoTe$_2$/*p$^+$*-CGS/Ni device at varying *p$^+$*-CGS BSF width in the span of 50-350 nm, doping volume of $10^{16}$-$10^{22}$ cm$^{-3}$ and volume defects of $10^{12}$-$10^{16}$ cm$^{-3}$. In Figure 6(a), the almost constant values of 0.97 V, 42.21 mA/cm$^2$, 82.73%, 32.92% are determined for the $V_{OC}$, $J_{SC}$, FF and PCE, one by one, when layer width is increased from 50 to 350 nm. Thinner width with high doping density of *p$^+$*-CGS compensates electronic and photonic recombination loss at the back absorber MoTe$_2$/Ni contact interface showing almost semi-metal characteristics.

In Figure 6(b), the $V_{OC}$, $J_{SC}$, FF, and PCE were insignificantly changed by varying doping concentration from $10^{16}$ to $10^{22}$ cm$^{-3}$ of the BSF under fixed values of parameters of the window layer, and absorption layers. The value of 0.96-0.97 V, 41.20-41.21 mA/cm$^2$, 81.32 – 82.83%, and 32.16 to 33.06 % were retained for the $V_{OC}$, $J_{SC}$, FF, and PCE, one by one at a varying doping density.

Similarly, the voltage, current, FF, and efficiency were found almost unchanged at 0.97 V, 42.21 mA/cm$^2$, 82.73%, and 32.92%, when the defect density is increased from $10^{12}$ c/m$^3$ to $10^{18}$ c/m$^3$ keeping fixed all the rest parameters related to window layer, absorption layer and BSF layer (Figure 6(c)). Formation of additional p-p junction at *p*-MoTe$_2$/*p$^+$*-CGS interface by the insertion of CGS BSF between MoTe$_2$ layer and Ni back contact. It originates ancillary electric field, which prevents the flow of electron and deflected electron, results in, improved $V_{OC}$, $J_{SC}$ with reduced reverse saturation (dark) current. Thus, the PCE with the BSF layer is enhanced from 31.53% to 32.92%.

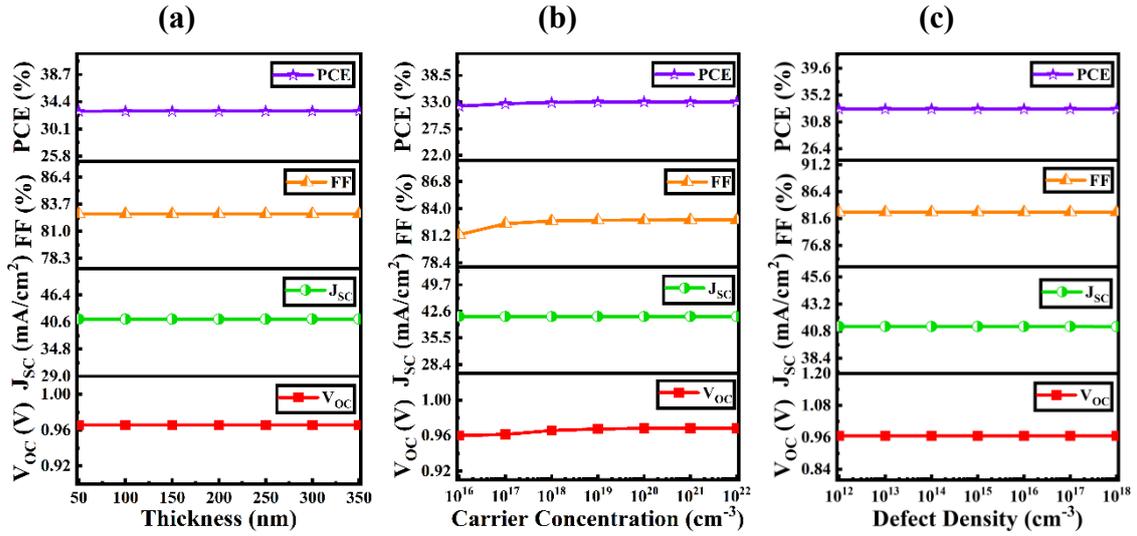

**Figure 6:** The change in PV parameters at varying (a) layer width, (b) carrier density, and (c) bulk defects of CGS portion of MoTe$_2$-based device.

### 3.3.2 Impact of BSF layer on MoTe$_2$ photovoltaic and photodetector

Figure 7 shows the variation in J-V, QE, R and D$^*$ characteristics with and without the BSF layer of $p^+$-CGS under fixed layer thickness 200 nm, the doping volume of $1.0\times10^{18}$ cm$^{-3}$, as well as defect volume of $1.0\times10^{15}$ cm$^{-3}$. In Figure 7(a), the 0.89 V and 41.19 mA/cm$^2$ is observed for the voltage and current without the addition of $p^+$-CGS BSF layer, which further improved to a 0.97 V and 41.21 mA/cm$^2$ with the inclusion of BSF in the middle of $p$-MoTe$_2$ and Ni contact which is used as back contact. This marked improvement in voltage and photocurrent signifies the role of the $p^+$-CGS BSF layer forming $n$-CdS/$p$-MoTe$_2$/$p^+$-CGS double heterostructure. On the other hand, in Figure 7(b), a noticeable enhancement in QE with improved absorption of incident photons is observed in the longer wavelength region from 700 to 1100 nm. This promotion in current is resulted owing to the BSF outcomes that reduces the recombination by mitigating the velocity of surface recombination which occurs from a high number of surface trapping centers in the boundary region [42].

From Figure 7(c), it is seen that, the peak responsivity of the photodetector reaches 0.73 A/W at wavelength of 1000 nm without the presence of the p$^+$-CGS layer. Moreover, with the incorporation of the BSF, the R grows slightly to 0.74 A/W at the wavelength of 1000 nm, that is in the NIR part. Notably, R remains elevated across the 900 to 1100 nm range, demonstrating the photosensor's effectiveness in the NIR region with the inclusion of the BSF layer.

Detectivity is a crucial parameter for photodetector devices, reflecting the capability to determine the weak optical sign and discern their execution in this regard [39]. Also, in Figure 7 (d), the peak detectivity without the CGS BSF is $6.00 \times 10^{15}$ Jones, measured at a wavelength of 1000 nm. With the addition of BSF, detectivity increases to $2.36 \times 10^{16}$ Jones at 1000 nm, within the near-infrared (NIR) region. This type of increment is happened because of the dependency of detectivity on dark current as seen in the equation 7, which in turn reduces owing to the higher $V_{OC}$ that results from higher built-in voltage. Notably, detectivity remains high across the 900 to 1100 nm range, highlighting the photosensor's effectiveness in the NIR section when the CGS layer is incorporated.

These facts have been happened because of the slightly increment on current as shown in Figure 7(a) that increases the QE as well as R and $D^*$. Because the R and $D^*$ are completely depend on QE as seen in equations 6 and 7.

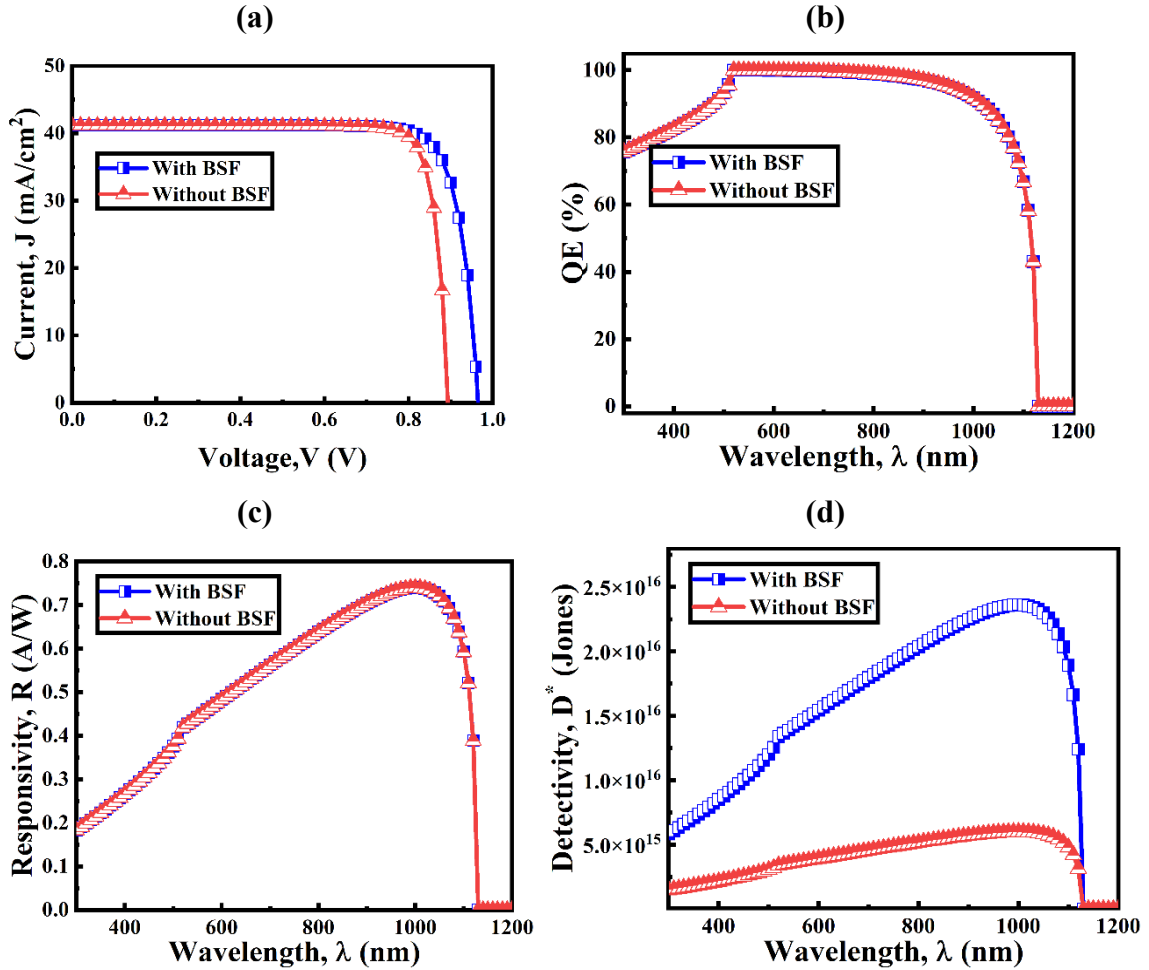

**Figure 7:** (a) J-V, (b) QE characteristics, (c) R and (d) $D^*$ graphs of $MoTe_2$-based double-heterojunction PD without and with the CGS layer.

**3.4 Device dependency on temperature and metal's work function**

Figure 8 displays the influence of operating temperature and work function of metals on the novel n-CdS/p-MoTe$_2$/p$^+$-CGS solar cell's PV parameters. The PV cell performance is noticeably affected by cell working Temperature. The PV performance varies with proportional to the charge carrier velocity and inversely proportional to the bond energy [43]. However, the reverse saturation current increased with decreasing bond energy [44]. Figure 8(a) shows the impact of temperature (WT) on PV parameters at 250-350 K. The V$_{OC}$ and FF decreased almost linearly from 1.00 V to 0.93 V and 85.24% with decreasing WT, eventually, the PCE markedly decreased from 35.06 to 30.73%, even though, the J$_{SC}$ was almost unchanged at a 41.21 mA/cm$^2$. But with the hike of temperature, the electronic energy gap is diminished which increases current fractionally and decreases the voltage pointedly with increased temperature because higher temperature generates more interface defects which increase series resistance and declines charge carrier diffusion length [45]. Our proposed solar device shows almost similar behavior. V$_{OC}$ decrease pointedly from 1.00 to 0.93 V and J$_{SC}$ hike a fraction from 41.06 to 41.34 mA/cm$^2$. These outcomes reveal the strong thermal stability of the newly designed ITO/*n*-CdS/*p*-MoTe$_2$/*p$^+$*-CGS/Ni device showing an efficiency of over 30%, even the WT over 350 K.

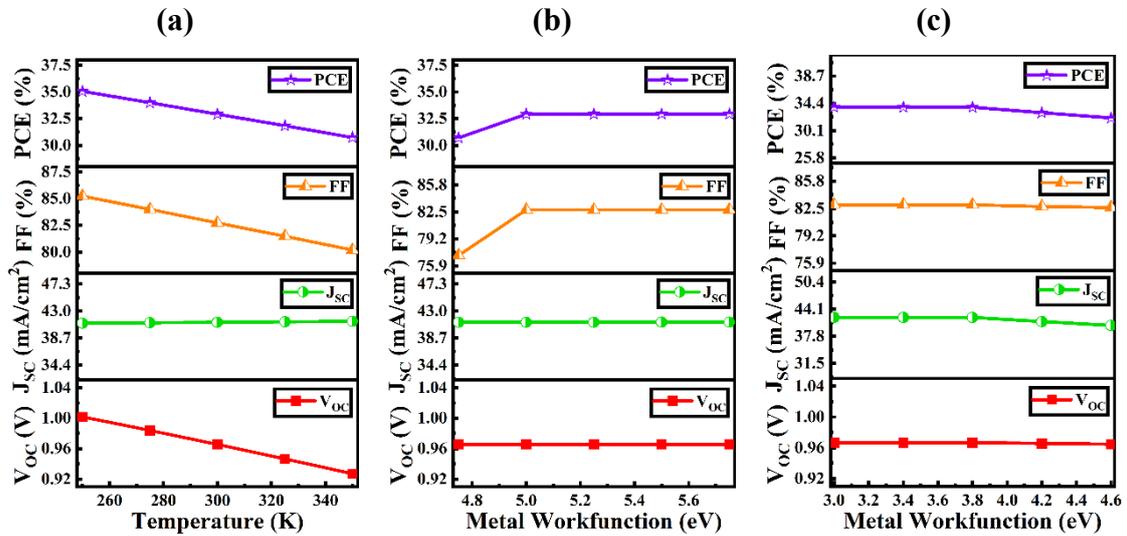

**Figure 8:** The dominance of the variation of (a) working temperature, (b) back and (c) front metal work function in PV characteristics of MoTe$_2$ PV cell.

Figure 8(b) and (c) exhibits the effect of metal work function of back and front metals, respectively on the performances of PV characteristics of this novel solar device. Aluminium (Al) and Nickel (Ni) have been embarked as front and rear metal contact layers with 4.2 eV and 5.2 eV work function (WF), in turn. These electrodes have been chosen considering the Flat Band energy of front and rear contacts. It is also seen from the figure that the effects of both front and back metal contacts are negligible.

### 3.5. Aggregate execution of this device

Table 2 demonstrates the summary of PV and PS parameters of MoTe$_2$ solar cell without and with the addition of a BSF portion forming a double junction heterostructure of $n$-CdS/$p$-MoTe$_2$/$p^+$-CGS. In a single-junction $n$-CdS/$p$-MoTe$_2$ solar cell and photosensor, the PCE is found to be 31.53% with V$_{OC}$ of 0.89 V, J$_{SC}$ of 41.19 mA/cm$^2$, FF of 85.58%, R of 0.74 A/W and D$^*$ of 6.00× 10$^{15}$ Jones which is further improved to 32.92% with 0.97 V, 41.21 mA/cm$^2$, 82.73%, 0.74 A/W and 2.36 × 10$^{16}$ Jones in turn, with additional $p^+$-CGS BSF as $n$-CdS/$p$-MoTe$_2$/$p^+$-CGS double-heterostructure photonic device. This significantly improved the PV performance with PCE over 32.92% incurs the strong potentiality and significance of the proposed device structure based on $n$-CdS window, $p$-MoTe$_2$, absorber, and $p^+$-CGS BSF layers.

**Table 2.** The photovoltaic frameworks of MoTe$_2$ device with and without the inclusion of $p^+$-CuGaSe$_2$ back surface layer.

| Limiting Factors | Device architecture | |
|---|---|---|
| | $n$-CdS/$p$-MoTe$_2$ | $n$-CdS/$p$-MoTe$_2$/$p^+$-CGS |
| J$_{SC}$ (mA/cm$^2$) | 41.19 | 41.21 |
| V$_{OC}$ (V) | 0.89 | 0.97 |
| FF (%) | 85.58 | 82.73 |
| PCE (%) | 31.53 | 32.92 |
| R (A/W) | 0.74 | 0.74 |
| D$^*$ (Jones) | 6.00× 10$^{15}$ | 2.36 × 10$^{16}$ |

### 4. Conclusions

The MoTe$_2$-based heterojunction SC and PS with a novel and potential $n$-CdS/$p$-MoTe$_2$/$p^+$-CGS layout have been designed and analyzed systematically by SCAPS-1D simulation software. The major optimized parameters such as layer thickness, doping and defects of the CdS, MoTe$_2$, and CGS parts have been determined initially. The optimized width of particular parts of 100, 700, and 200 nm, carrier of 1.0 × 10$^{18}$, 1.0 × 10$^{15}$, and $1.0 \times 10^{18}$ cm$^{-3}$, and bulk defect volume of $1.0 \times 10^{15}$, $1.0 \times 10^{15}$, and $1.0 \times 10^{15}$ cm$^{-3}$ for CdS, MoTe$_2$ and CGS, respectively have been obtained. Subsequently, the markedly improved PCE of 32.92% has been obtained in the proposed device structure with $p^+$-CGS BSF (which was 31.53% without BSF) along with a V$_{OC}$ of 0.97 V, J$_{SC}$ of 41.21 mA/cm$^2$, and FF of 82.73%. Also, the photosensor characteristics such as responsivity and detectivity have been obtained as 0.74 A/W and $2.36 \times 10^{16}$ Jones, in turn. These detailed studies and outcomes reveal the strong potentiality of MoTe$_2$ absorber, with CdS as window as well as CGS as BSF layers, which has not been introduced before and provide significant resources for practically fabricating the MoTe$_2$-based environment-friendly, cost-effective, and high-efficiency photonic devices.

**Acknowledgments**

The authors highly appreciate Dr. Marc Burgelman, University of Gent, Belgium, for providing SCAPS simulation software.

*****Corresponding author:** E-mail: jak_apee@ru.ac.bd (Jaker Hossain)

**Data availability:** Simulation details and associated data are available free of charge from authors upon reasonable request.

**Disclosure:** The authors declare no competing financial interest.

**Declaration of generative AI and AI-assisted technologies:** None of the authors use any AI or AI-assisted technologies in writing this manuscript.